\documentclass[letterpaper,twocolumn,prl,aps,showpacs,superscriptaddress,floatfix]{revtex4-1}
\usepackage[latin1]{inputenc}
\usepackage{bm}
\usepackage[usenames]{color}
\usepackage{multirow}
\usepackage{amssymb}
\usepackage{amsbsy}
\usepackage{amsmath}
\usepackage{stmaryrd}
\usepackage{graphicx}
\usepackage{epsfig}
\usepackage{placeins}
\usepackage{ulem}
\makeatletter

\begin{document}

\title{Spin-Current Autocorrelations from Single Pure-State Propagation}

\author{Robin Steinigeweg}
\email{r.steinigeweg@tu-bs.de} \affiliation{Institute for Theoretical
Physics, Technical University Braunschweig, D-38106 Braunschweig,
Germany}

\author{Jochen Gemmer}
\email{jgemmer@uos.de} \affiliation{Department of Physics,
University of Osnabr\"uck, D-49069 Osnabr\"uck, Germany}

\author{Wolfram Brenig}
\email{w.brenig@tu-bs.de} \affiliation{Institute for Theoretical
Physics, Technical University Braunschweig, D-38106 Braunschweig,
Germany}

\date{\today}

\begin{abstract}
We demonstrate that the concept of quantum typicality allows for significant
progress in the study of real-time spin dynamics and transport in quantum
magnets. To this end, we present a numerical analysis of the spin-current
autocorrelation function of the antiferromagnetic and anisotropic spin-$1/2$
Heisenberg chain as inferred from propagating only a {\it single} pure state,
randomly chosen as a ``typical'' representative of the statistical
ensemble. Comparing with existing time-dependent density-matrix renormalization
group (tDMRG) data, we show that typicality is fulfilled extremely well,
consistent with an error of our approach, which is perfectly under control and
vanishes in the thermodynamic limit. In the long-time limit, our results provide
for a new benchmark for the enigmatic spin Drude weight, which we obtain from
chains as long as $L=33$ sites, i.e., from Hilbert spaces of dimensions almost
${\cal O}(10^4)$ larger than in existing exact-diagonalization studies.
\end{abstract}


\maketitle

{\it Introduction}. Understanding relaxation and transport dynamics
in quantum many-body systems is one of the most ambitious aims of
condensed-matter physics and is experiencing an upsurge of interest
in recent years, both experimentally and theoretically. On the one
hand, the advent of ultracold atomic gases raises challenging
questions about the thermalization of isolated many-body systems
\cite{cazalilla2010}. On the other hand, future information technologies
such as spintronics call for a deeper insight into transport processes of quantum
degrees of freedom such as spin excitations. While spin transport in conventional
nano-systems \cite{appelbaum2007, tombros2007, stern2008, kuemmeth2008} is
inevitably linked to itinerant charge-carrier dynamics, Mott-insulating
quantum magnets allow for pure spin currents, opening new perspectives
in quantum transport. Magnetic transport in one-dimensional (1D) quantum
magnets has attracted considerable attention in the past decade due to the
discovery of very large magnetic heat-conduction \cite{sologubenko2000,
hess2001, hlubek2010} and long nuclear magnetic relaxation times
\cite{thurber2001, kuehne2010}; so far, however, pure spin transport
remains to be observed experimentally.

Within the large body of theoretical work accumulated
\cite{shastry1990, zotos1999, benz2005, prosen2011, prosen2013,
narozhny1998, heidrichmeisner2003, herbrych2011,
steinigeweg2009, steinigeweg2011, prelovsek2013, mierzejewski2010,
steinigeweg2012-1, alvarez2002, grossjohann2010, langer2009,
karrasch2012, karrasch2013, karrasch2014, znidaric2011,
steinigeweg2012-2, sirker2009, fujimoto2003}, the dissipation
of magnetization currents is a key issue. This issue has been studied
extensively at zero momentum and frequency in connection with the well-known
spin Drude weight: It is the non-decaying contribution of the spin current and
signals dissipationless transport. A paradigmatic model in these
studies is the Heisenberg chain with periodic boundary conditions
($\hbar = 1$),
\begin{equation}
H = J \sum_{r=1}^L (S_r^x S_{r+1}^x + S_r^y S_{r+1}^y + \Delta \,
S_r^z S_{r+1}^z) \, ,
\end{equation}
where $S_r^{x,y,z}$ are the components of spin-$1/2$ operators
at site $r$. $L$ is the total number of sites, $J > 0$ the
antiferromagnetic exchange coupling constant, and $\Delta$ the
anisotropy. 

At zero temperature, $T = 0$, early work \cite{shastry1990} showed
that the Drude weight is finite in the gapless regime $\Delta \leq 1$
(metal) but vanishes in the gapped case $\Delta > 1$ (insulator). At
finite temperature, $T > 0$, Bethe-Ansatz results \cite{zotos1999,
benz2005} support a qualitatively similar picture, but with a
disagreement at the isotropic point $\Delta=1$. Recent progress in
combining quasi-local conservation laws and the Mazur's inequality
has lead to a rigorous lower bound to the Drude weight at high
temperatures $T \gg J$ \cite{prosen2011, prosen2013}, which is very
close to the Bethe-Ansatz solution but still allows for a vanishing
Drude weight at $\Delta = 1$.

Numerically, a large variety of sophisticated methods has been applied
to spin transport in Heisenberg chains, including full exact diagonalization
(ED) \cite{narozhny1998, heidrichmeisner2003, herbrych2011,
steinigeweg2009, steinigeweg2011}, $T >0$ Lanczos \cite{prelovsek2013}, quantum
Monte-Carlo \cite{alvarez2002, grossjohann2010},
wave-packet propagation by tDMRG \cite{langer2009}, real-time
correlator calculations \cite{karrasch2012, karrasch2013, karrasch2014}, and
Lindblad quantum master equations \cite{znidaric2011}. The Drude
weight, however, is only available from ED and tDMRG. Since, as of today, ED is
restricted to chains of length $L \sim 20$, the long-time (low-frequency)
regime is still governed by finite-size effects and intricate extrapolation
schemes to the thermodynamic limit have been invoked, with different results
depending on details -- including remarkable differences between even and
odd $L$ \cite{herbrych2011}, or grand-canonical and canonical ensembles
\cite{karrasch2013}. Alternatively, tDMRG is exceedingly more powerful
w.r.t.~system size and chains with $L \sim 200$ are accessible; however,
the method is still confined to a maximum time scale depending on $\Delta$
\cite{karrasch2012, karrasch2013, karrasch2014}, with an ongoing progress
to increase this scale \cite{note}. As of today, the latter scale is too
short for a reliable extraction of the Drude weight at the isotropic point
$\Delta = 1$ \cite{karrasch2013}.

Therefore, no consistent picture on the Drude weight for $T \neq 0$ is
available at $\Delta \sim 1$. It is worth mentioning that, apart from
Drude weights indicating ballistic dynamics, steady-state bath scenarios
\cite{znidaric2011} and classical simulations \cite{steinigeweg2012-2}
suggest super-diffusive dynamics at $T \gg J$, while bosonization predicts
diffusion at sufficiently low temperatures $J \gg T > 0$ \cite{sirker2009}.

In this situation, our Letter takes a fresh perspective, rooted in
quantum statistical physics and leading to a surprisingly simple,
yet very powerful numerical approach to evaluate finite-temperature
time-dependent correlation functions in general and the dynamics of
spin currents in the Heisenberg chain in particular. This approach
is intimately related to the emergent concept of quantum typicality
\cite{gemmer2003, goldstein2006, popescu2006, reimann2007, white2009,
bartsch2009, sugiura2012, elsayed2013} and overcomes the
restriction of ED to small system sizes without a restriction to short
times. Specifically, in this Letter, we will unveil that {\it(i)} quantum
typicality is fulfilled extremely well for system sizes $L \sim 30$ down
to temperatures $T/J = 0.5$ and that {\it(ii)} our approach yields exact
information on an extended time window in the thermodynamic limit. Furthermore,
because our approach is not restricted to short times, we are able to {\it(iii)}
calculate the Drude weight for chains as long as $L=33$ sites, i.e., for
Hilbert spaces almost ${\cal O}(10^4)$ times larger than in present ED at
$L \leq 20$. Thus, our results {\it (iv)} severely constrain any remaining
speculations on the long-standing issue of the finite-temperature spin Drude
weight of the isotropic Heisenberg chain.

{\it Numerical approach.} The well-known spin-current operator
$j = J \sum_r (S_r^x S_{r+1}^y - S_r^y S_{r+1}^x)$ follows from the
continuity equation \cite{heidrichmeisner2003}. Within the framework
of linear response theory, we are interested in the autocorrelation
function at inverse temperatures $\beta = 1/T$ ($k_B = 1$),
\begin{equation}
C(t) = \frac{\langle j(t) \, j \rangle}{L} = \frac{\text{Tr}
\{e^{-\beta H} j(t) \, j \}}{L \, \text{Tr}\{e^{-\beta H}\}}
\, , \label{exact}
\end{equation}
where the time argument of $j$ has to be understood w.r.t.\ the
Heisenberg picture, $j = j(0)$, and $C(0) = J^2/8$ at high
temperatures $\beta \to 0$. For the validity of linear response
theory, see Refs.~\onlinecite{mierzejewski2010,steinigeweg2012-1}.

% ------------------------------- FIGURE 1 ----------------------------------
\begin{figure}[t]
\includegraphics[width=1.0\columnwidth]{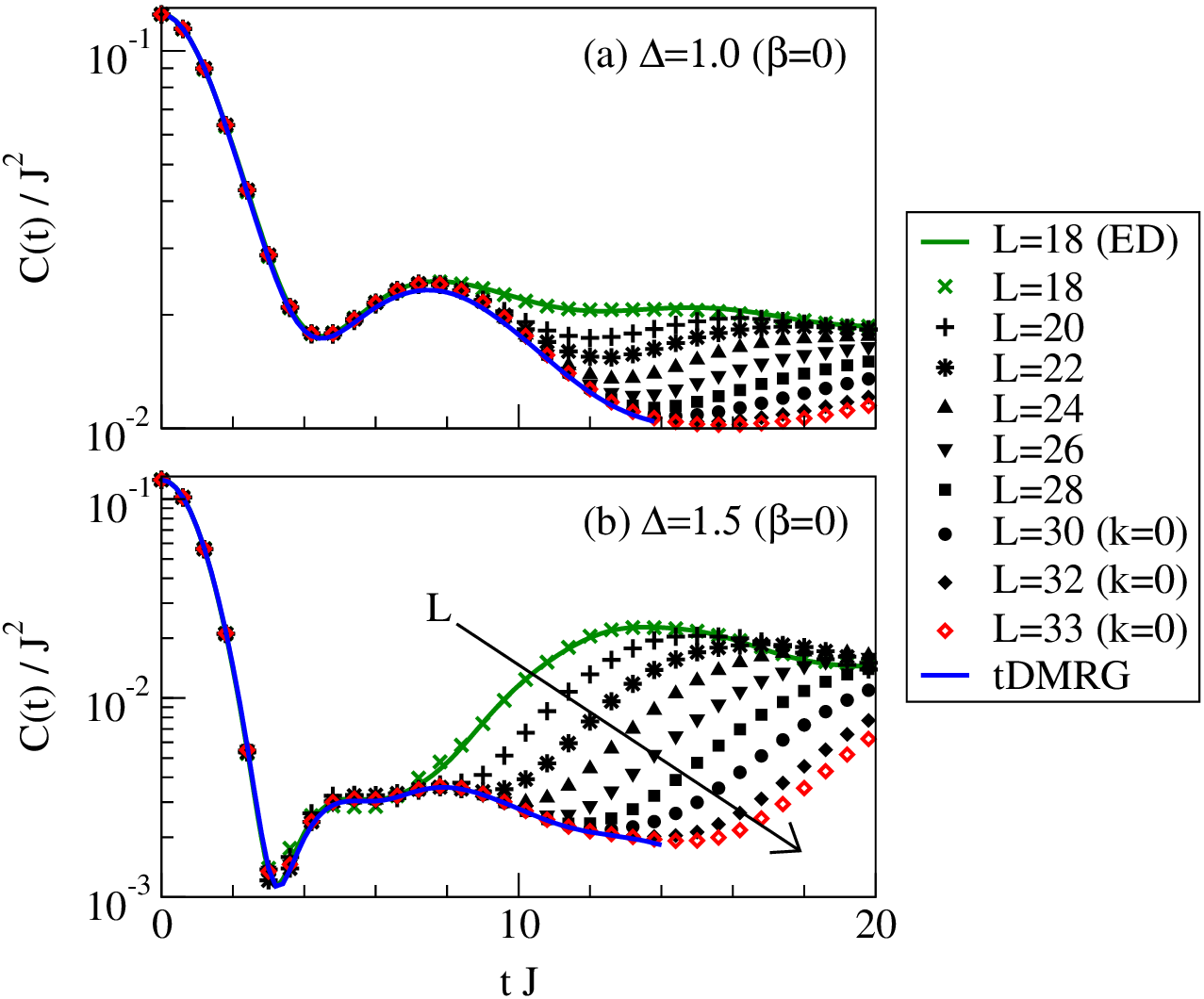}
\caption{(color online) Spin-current autocorrelation function
$C(t)$ at $\beta \to 0$ for (a) $\Delta = 1.0$ and (b) $\Delta = 1.5$,
numerically obtained for $L =18$ using the full statistical ensemble
(green curve) and larger $L \geq 18$ using a single pure state (symbols),
shown in a semi-log plot. The very high accuracy is illustrated by
comparing to available tDMRG data for $L=200$ \cite{karrasch2012,
karrasch2014} (blue curve).}
\label{Fig1}
\end{figure}
% ---------------------------------------------------------------------------

The basic idea underlying our numerical approach is to replace the
trace $\text{Tr}\{\bullet\}=\sum_n \langle n | \bullet | n \rangle$
by a scalar product involving a single pure state $| \psi \rangle$
drawn at random. More precisely, following the concept of quantum
typicality \cite{gemmer2003, goldstein2006, reimann2007, popescu2006},
$| \psi \rangle$ is drawn at random from a probability distribution
which is invariant under all possible unitary transformations in
Hilbert space (Haar measure \cite{bartsch2009}). Drawing such a
random pure state $| \psi \rangle$ and abbreviating $| \psi_\beta
\rangle = e^{-\beta H/2} | \psi \rangle$, we can approximate the
autocorrelation function by \cite{bartsch2009, sugiura2012, elsayed2013}
\begin{equation}
C(t) = \frac{\langle \psi_\beta | j(t) \, j | \psi_\beta \rangle}{L
\, \langle \psi_\beta | \psi_\beta \rangle} + {\cal O} \Big
(\frac{\sqrt{\langle j(t) \, j \, j(t) \, j \rangle}} {L \,
\sqrt{d_\text{eff}}} \Big) \, , \label{approximative}
\end{equation}
where the second term is a random variable with zero mean and
standard deviation $\propto 1/\sqrt{d_\text{eff}}$, with
$d_\text{eff}$ as the effective dimension of the Hilbert space.
In the limit of high temperatures $\beta \to 0$, $d_\text{eff}
= 2^L$. Hence, if the chain length $L$ is increased, the error
decreases exponentially with $L$ and the approximation becomes
accurate rather quickly. At arbitrary temperatures, $d_\text{eff}
= \text{Tr} \{ e^{-\beta(H-E_0)} \}$ is the partition function
and $E_0$ the ground-state energy. This expression also scales
exponentially with $L$ \cite{steinigeweg2013}, rendering the
approximation exact once again for $L \rightarrow \infty$,
however less quickly. In any case, the error in
Eq.~(\ref{approximative}) can be reduced additionally by
averaging over several random pure states $|\psi \rangle$.

% ------------------------------- FIGURE 2 ----------------------------------
\begin{figure}[t]
\includegraphics[width=0.8\columnwidth]{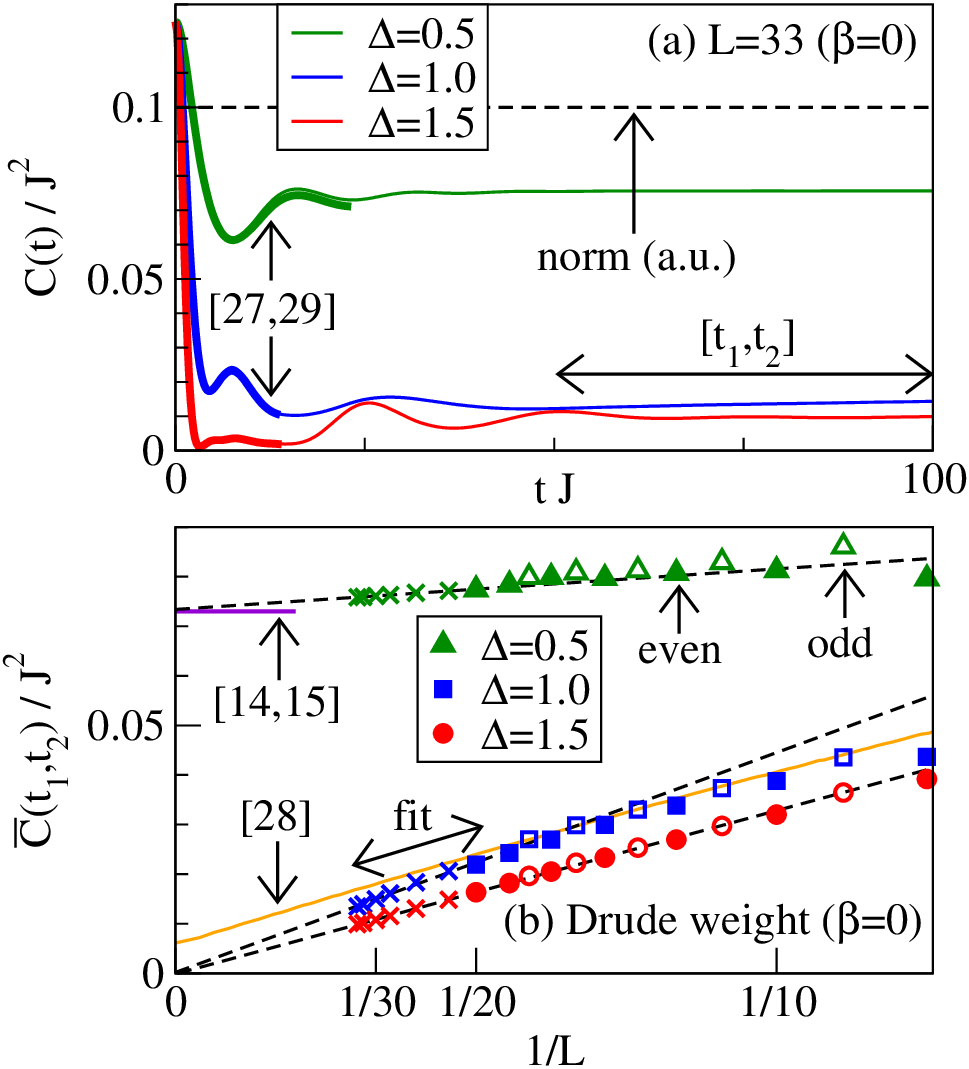}
\caption{(color online) (a) Long-time limit of the spin-current
autocorrelation function $C(t)$ at $\beta \to 0$ for $\Delta = 0.5$,
$1.0$, and $1.5$, numerically obtained using a single pure state
(thin solid curves). The well-conserved norm is indicated (thin
dashed curve). Available tDMRG data for $L=200$
\cite{karrasch2012, karrasch2014} are shown (thick solid curves). (b)
Finite-size scaling of the spin Drude weight, extracted in the time
interval $[t_1 \, J, t_2 \, J] = [50,100]$ (closed [open] symbols
for even [odd] $L$; $L \leq 20$: full statistical ensemble; and $L \gg 20$:
single pure state). Simple $1/L$ fits to large $20 \leq L \leq 33$ are
depicted (dashed lines), and at $\Delta = 1.0$ the odd-site fit to $L
\leq 19$ performed in Ref.~\onlinecite{karrasch2013} (solid curve). At
small $\Delta = 0.5$, T.~Prosen's strict analytic lower bound \cite{prosen2011,
prosen2013} is indicated (horizontal line).}
\label{Fig2}
\end{figure}
% ---------------------------------------------------------------------------

The central advantage of Eq.~(\ref{approximative}) is that the
first term on the r.h.s.\ can be evaluated without diagonalization
of the Hamiltonian. This can be seen by introducing two pure states:
The first is $| \Phi_\beta(t) \rangle = e^{-\imath H t -\beta H/2} \,
| \psi \rangle$ and the second is $| \varphi_\beta(t) \rangle =
e^{-\imath H t} \, j \, e^{-\beta H/2} \, |\psi \rangle$. Then,
\begin{equation}
\langle \psi_\beta | j(t) \, j | \psi_\beta \rangle = \langle
\Phi_\beta(t) | j | \varphi_\beta(t) \rangle \, .
\end{equation}
The $t$ ($\beta$) dependence of the two states can be calculated
by iterating in real (imaginary) time using, e.g., Runge-Kutta
\cite{elsayed2013, steinigeweg2013} or Chebyshev \cite{deraedt2007,
jin2010} schemes. We find that a fourth order Runge-Kutta (RK4)
scheme with a small discrete time step $\delta t \, J = 0.01 \ll 1$
is sufficient. Although the algorithm does not require to
save Hamiltonians and observables in memory, it is convenient
w.r.t.\ run time to do so, requiring only sparse matrices with $L \,
2^L \ll 2^{2L}$ elements. It is further convenient to choose the
random pure state $| \psi \rangle$ from the common eigenbasis of
symmetries, e.g., translation invariance and rotations about the
$z$ axis, taking full advantage of the block structure of the
Hamiltonian in that basis. In this way, we are able to treat chains
as long as $L=33$, where the full Hilbert-space dimension is
huge: $2^{33} \approx 10^{10}$. (In that case the dimension of the
largest symmetry subspace is $3.5 \cdot 10^7$.) It is worth mentioning
that symmetries, in combination with massive parallelization on super
computers \cite{jin2010}, have the potential to reach $L \sim 40$
in the future.

{\it Results.} Let us begin with the high-temperature limit
$\beta \to 0$ and systems of intermediate size $L = 18$. In
Fig.~\ref{Fig1} (a) we first compare the exact autocorrelation
function in Eq.~(\ref{exact}), obtained from ED, at the isotropic
point $\Delta = 1$ with the approximation in Eq.~(\ref{approximative}),
obtained from RK4, using a single pure random state $| \psi \rangle$.
At all times, the agreement between Eq.~(\ref{exact}) and
Eq.~(\ref{approximative}) is remarkably good. This agreement is
underlined even more through our usage of a log $y$ axis, emphasizing
relative rather than only absolute accuracy. In view of this agreement,
and with any remaining error decreasing exponentially with $L$, we
can safely consider the Eq.~(\ref{approximative}) as almost exact
for $L > 18$ and we will neglect any averaging over random pure
states $| \psi \rangle$. By increasing $L$ in Fig.~\ref{Fig1} (a),
we show that the curve of the autocorrelation function gradually
converges in time towards the thermodynamic limit. For the maximum
$L=33$ the curve is converged up to times $t \, J \sim 15$ with no
visible finite-size effects in the semi-log plot. Note that for the
three largest $L \geq 30$ we restrict ourselves to a single translation
subspace to reduce computational effort at high temperatures $\beta
\to 0$, see also Ref.~\cite{herbrych2011}. Next, we compare to existing
tDMRG data for $L = 200$ \cite{karrasch2012}. It is intriguing to see
that our results agree up to very high precession. On the one hand, this
observation is the most convincing demonstration of quantum typicality
so far. On the other hand, it indicates that our approach yields exact
information on an extended time window in the thermodynamic limit. These
latter points are two main results of this Letter.

% ------------------------------- FIGURE 3 ----------------------------------
\begin{figure}[t]
\includegraphics[width=0.8\columnwidth]{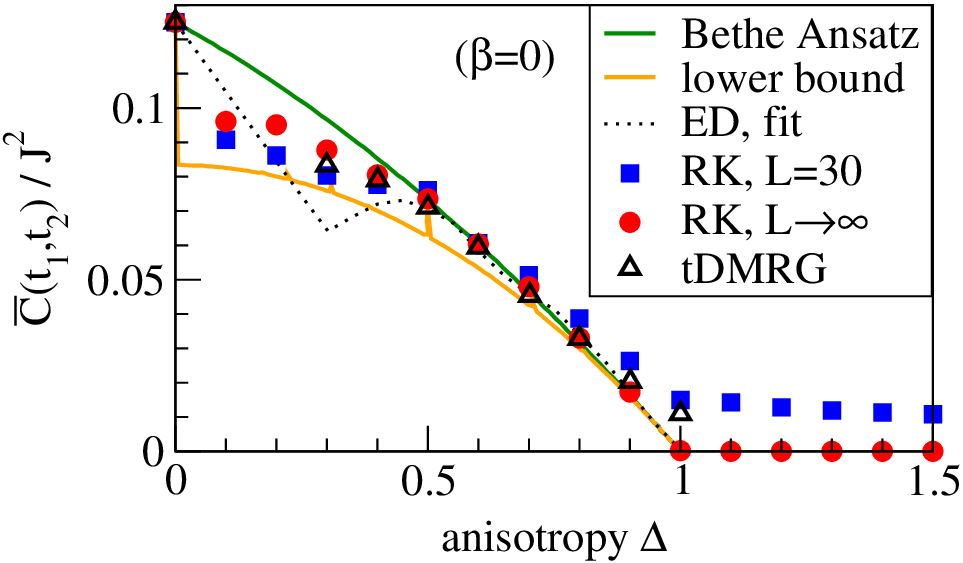}
\caption{(color online) High-temperature Drude weight $\bar{C}$
from single pure-state propagation w.r.t.\ the anisotropy $\Delta$
(closed symbols), compared to the thermodynamic Bethe-Ansatz
\cite{zotos1999}, T.~Prosen's strict analytic lower bound
\cite{prosen2011, prosen2013}, the fit to ED at zero magnetization
and odd sites performed in Ref.~\onlinecite{herbrych2011}, and
tDMRG \cite{karrasch2012} (see also Ref.~\onlinecite{karrasch2013}
for a different point of view on the tDMRG data point at $\Delta
= 1$).}
\label{Fig3}
\end{figure}
% ---------------------------------------------------------------------------

In Fig.~\ref{Fig1} (b) we show a second calculation for a larger anisotropy
$\Delta = 1.5$. Clearly, $C(t)$ decays to almost zero rapidly. However, a small
long-time tail remains. This tail has been anticipated already on the basis
of ED at $L=20$ \cite{steinigeweg2009}, leading to a positive correction to
the diffusion constant from perturbation theory \cite{steinigeweg2011}. Note
that the tail is not connected to the Drude weight \cite{karrasch2014}, as
discussed in more detail later.

Next we study the long-time limit. In Fig.~\ref{Fig2} (a) we show $C(t)$
for $\beta \to 0$ and $\Delta = 0.5$, $1.0$, and $1.5$. We also depict the
norm of $| \Phi_\beta(t) \rangle$, which is practically constant, as $|
\varphi_\beta(t) \rangle$ is also. This clearly demonstrates that the RK4
scheme works properly at such long times. This figure also proves {\it(i)}
the saturation of $C(t)$ at rather long times $t J \sim 50$ and that {\it(ii)}
we can hardly infer the saturation value from our short-time data in
Fig.~\ref{Fig1}.

Because the long-time limit is governed by finite-size effects, we are now
going to perform a proper finite-size scaling in terms of the Drude weight.
To this end, let us define the Drude weight $\bar{C} = 1/(t_2-t_1) \int_{t_1}^{t_2}
\! C(t)\, \mathrm{d}t$ as the average over the time interval $[t_1 \, J, t_2 \, J]
= [50,100]$. The Drude weight has been extracted similarly in
Ref.~\onlinecite{steinigeweg2009}, yielding the correct zero-frequency value
\cite{heidrichmeisner2003}.

% ------------------------------- FIGURE 4 ----------------------------------
\begin{figure}[t]
\includegraphics[width=1.0\columnwidth]{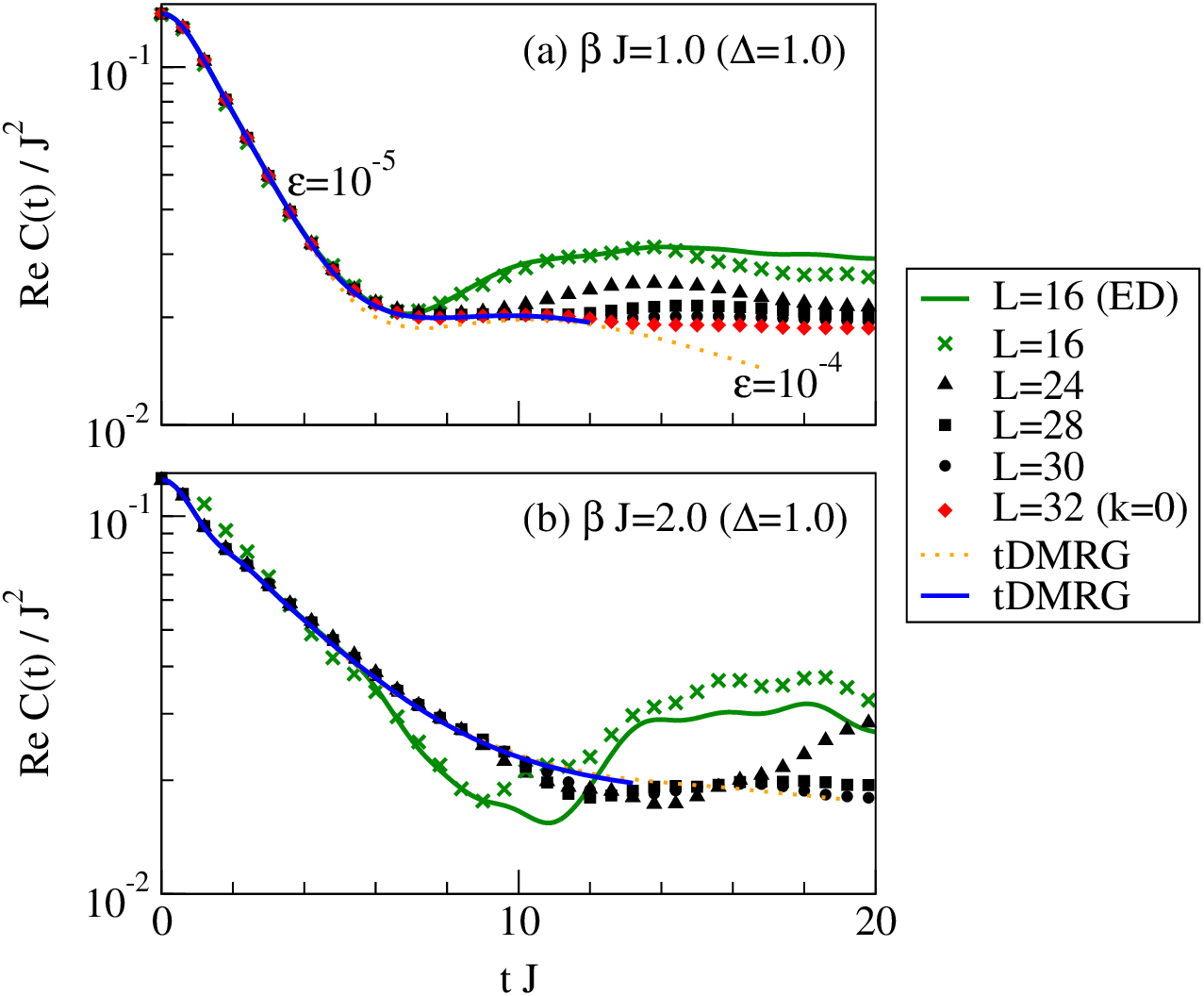}
\caption{(color online) Spin-current autocorrelation function $C(t)$
for $\Delta = 1.0$ at (a) $\beta \, J = 1.0$ and (b) $\beta \, J = 2.0$,
numerically obtained for $L =16$ using the full statistical ensemble
(green curve) and larger $L \geq 16$ using a single pure state (symbols),
shown in a semi-log plot. Available tDMRG data for $L=200$ \cite{karrasch2012,
karrasch2013} are depicted for two values of the discarded weight $\epsilon$
(blue and orange curve).}
\label{Fig4}
\end{figure}
% ---------------------------------------------------------------------------

In Fig.~\ref{Fig2} (b) we depict the resulting Drude weight vs.\ the
inverse length $1/L$ for anisotropies $\Delta = 0.5$, $1.0$, and $1.5$.
While for $L > 20$ we extract the Drude weight from the approximation in
Eq.~(\ref{approximative}) (denoted by crosses), we use the exact expression
in Eq.~(\ref{exact}) for $L \leq 20$ (denoted by other symbols), to avoid
typicality errors at small $L$. We also indicate the results of $1/L$ fits,
solely based on data points for $L \geq 20$, to avoid the influence of even-odd
effects and the need of $(1/L)^{i>1}$ corrections at small $L$. Remarkably,
for $\Delta = 0.5$, the resulting fit is close to {\it all} data points.
Extracting the thermodynamic limit $L \to \infty$ from the fits, we find a
non-zero Drude weight in convincing agreement with the rigorous lower bound
of Refs.~\onlinecite{prosen2011, prosen2013}. The situation is rather similar
for $\Delta = 1.5$ but with a vanishing Drude weight for $L \to \infty$,
consistent with previous work \cite{heidrichmeisner2003}. Certainly, the
isotropic point $\Delta = 1.0$ is most interesting. Here, the $L \geq 20$ fit
is not close to that obtained from only small $L < 20$. In fact, the
extrapolation yields much smaller values for the Drude weight than the
finite values suggested in previous works, based on either smaller $L$
\cite{heidrichmeisner2003, karrasch2013} or shorter $t$ \cite{karrasch2012}
(see also Ref.~\onlinecite{karrasch2013} for a detailed discussion). In
fact, our approach is consistent with a vanishing Drude weight
for $L \to \infty$. This is another main result of this Letter.

In Fig.~\ref{Fig3} we summarize the finite-size values for the Drude
weight for fixed $L=30$ and various anisotropies $0 \leq \Delta \leq 1.5$.
Additionally, we indicate the extrapolated values for $L \to \infty$
using fits to, e.g., only even sites, which is important closer
to $\Delta = 0$. Remarkably, all extrapolated values lie above the
rigorous lower bound of Refs.~\onlinecite{prosen2011, prosen2013} and,
in the anisotropy range $0.4 \lesssim \Delta \leq 1.5$, also agree with
the Bethe-Ansatz solution of Ref.~\cite{zotos1999}. They further agree with
an alternative extrapolation on the basis of small $L$ \cite{herbrych2011},
using a different statistical ensemble and only odd sites. In the
{\it vicinity} of $\Delta = 0$, where relaxation is slow, our definition
of the Drude weight in terms of $\bar{C}$ may include low-frequency
contributions. Still we lie above the lower bound but observe deviations
from the Bethe-Ansatz solution. These deviations are well-known to occur
in numerical studies using finite systems \cite{herbrych2011}, due to the
very high degeneracy at $\Delta = 0$.

Finally, we turn to finite temperatures $\beta \neq 0$. Clearly, our numerical
approach has to break down for $\beta \to \infty$, i.e., $T \to 0$, due to the
reduction of the effective Hilbert space dimension in Eq.~(\ref{approximative}).
Moreover, also the exact expression in Eq.~(\ref{exact}) is governed by large
finite-size effects for $\beta \, J \gg 2$, at least at $L \sim 30$
\cite{prelovsek2013}. Thus, for a numerical approach to $L \sim 30$, reasonable
temperatures are $\beta \, J \sim 2$. For this range of $\beta$ the approximation
is still justified and averaging is crucial for $\beta \gg 2$ only.

In Fig.~\ref{Fig4} (a) and (b) we compare the approximation in
Eq.~(\ref{approximative}), calculated by RK4, and the exact expression in
Eq.~(\ref{exact}), calculated by ED, for a system of intermediate size $L=16$
and the two lower temperatures $\beta \, J = 1$ and $2$, with the focus on
$\Delta = 1$. While deviations appear at $\beta \, J = 2$, these deviations
manifest as random fluctuations rather than systematic drifts and may be
compensated by additional averaging over several pure states. However, one
can expect that these deviations disappear for significantly larger $L$.
Again, we prove this by comparing with existing tDMRG data for $L=200$
\cite{karrasch2012, karrasch2013}. This result illustrates the power of
our numerical approach at finite temperatures. Moreover, taking into account
{\it(i)} the simple structure of the curve, {\it(ii)} the semi-log plot,
and {\it(iii)} the combination of tDMRG with our numerical approach,
Fig.~\ref{Fig4} (a) is very indicative to non-zero Drude weights at $\beta
\neq 0$, in contrast to $\beta \to 0$, which is another main result of
this Letter.

{\it Conclusion.} We used the emergent concept of quantum typicality to
obtain an alternative and innovative numerical approach to several timely
issues regarding spin transport in anisotropic and antiferromagnetic
spin-$1/2$ Heisenberg chains. We showed that quantum typicality is fulfilled
extremely well for system sizes $L \sim 30$ down to temperatures $T/J = 0.5$.
Because our approach is not restricted to short times, we were able to
calculate the Drude weight for chains as long as $L=33$ sites. This enabled
us to drastically narrow down any options for the long-standing question of
Drude weights in the isotropic Heisenberg chain, which we find to be very
small or vanishing at high temperatures. We hope that in the future our
approach complements other numerical techniques in a much broader
context, including problems with few symmetries \cite{karahalios2009}
and/or in two dimensions \cite{chaloupka2010}. In non-integrable systems
we expect very small finite-size effects for the huge Hilbert spaces
reachable by our approach.

We sincerely thank H. Niemeyer, P.~Prelov\v{s}ek, and J. Herbrych
for fruitful discussions as well as C. Karrasch and F. Heidrich-Meisner
for tDMRG data and helpful comments. Part of this work has been done at
the Platform for Superconductivity and Magnetism, Dresden (W.B.). Part
of this work has been supported by DFG FOR912 Grant No.~BR 1084/6-2 and
EU MC-ITN LOTHERM Grant No.~PITN-GA-2009-238475.

\newpage


\begin{thebibliography}{99}

\bibitem{cazalilla2010}
M. A. Cazalilla and M. Rigol, New J. Phys. {\bf 12}, 055006 (2010);
and references therein.

\bibitem{appelbaum2007}
I. Appelbaum, B. Huang, and D. J. Monsma,
Nature {\bf 447}, 295 (2007).

\bibitem{tombros2007}
N. Tombros {\it et al.},
Nature {\bf 448}, 571 (2007).

\bibitem{stern2008}
N. P. Stern {\it et al.},
Nature Phys. {\bf 4}, 843 (2008).

\bibitem{kuemmeth2008}
F. Kuemmeth {\it et al.},
Nature {\bf 452}, 448 (2008).

\bibitem{sologubenko2000} A. V. Sologubenko {\it et al.},
Phys. Rev. Lett. {\bf 84}, 2714 (2000).

\bibitem{hess2001} C. Hess {\it et al.},
Phys. Rev. B {\bf 64}, 184305 (2001).

\bibitem{hlubek2010} N. Hlubek {\it et al.},
Phys. Rev. B {\bf 81}, 20405R (2010).

\bibitem{thurber2001} K. R. Thurber {\it et al.},
Phys. Rev. Lett {\bf 87}, 247202 (2001).

\bibitem{kuehne2010} H. K\"{u}hne {\it et al.},
Phys. Rev. B {\bf 80}, 045110 (2009).

\bibitem{shastry1990} B. S. Shastry and B. Sutherland,
Phys. Rev. Lett. {\bf 65}, 243 (1990).

\bibitem{zotos1999} X. Zotos,
Phys. Rev. Lett. {\bf 82}, 1764 (1999).

\bibitem{benz2005} J. Benz {\it et al.},
J. Phys. Soc. Jpn. {\bf 74}, 181 (2005).

\bibitem{prosen2011} T. Prosen,
Phys. Rev. Lett. {\bf 106}, 217206 (2011).

\bibitem{prosen2013} T. Prosen and E. Ilievski,
Phys. Rev. Lett. {\bf 111}, 057203 (2013).

\bibitem{narozhny1998} B. N. Narozhny, A. J. Millis, and N. Andrei,
Phys. Rev. B {\bf 58}, 2921R (1998).

\bibitem{heidrichmeisner2003} F. Heidrich-Meisner {\it et al.},
Phys. Rev. B {\bf 68}, 134436 (2003).

\bibitem{herbrych2011} J. Herbrych, P. Prelov\v{s}ek, and X. Zotos,
Phys. Rev. B {\bf 84}, 155125 (2011).

\bibitem{steinigeweg2009} R. Steinigeweg and J. Gemmer,
Phys. Rev. B {\bf 80}, 184402 (2009).

\bibitem{steinigeweg2011} R. Steinigeweg and W. Brenig,
Phys. Rev. Lett. {\bf 107}, 250602 (2011).

\bibitem{prelovsek2013} A recent review is given in: P. Prelov\v{s}ek and J. Bon\v{c}a,
{\it Ground State and Finite Temperature Lanczos Methods} in
{\it Strongly Correlated Systems},
Solid-State Sciences {\bf 176} (Springer, Berlin, 2013).

\bibitem{mierzejewski2010} M. Mierzejewski and P. Prelov\v{s}ek,
Phys. Rev. Lett. {\bf 105}, 186405 (2010).

\bibitem{steinigeweg2012-1} R. Steinigeweg {\it et al.},
Phys. Rev. B {\bf 85}, 214409 (2012).

\bibitem{alvarez2002} J. V. Alvarez and C. Gros,
Phys. Rev. Lett. {\bf 88}, 077203 (2002).

\bibitem{grossjohann2010} S. Grossjohann and W. Brenig,
Phys. Rev. B {\bf 81}, 012404 (2010).

\bibitem{langer2009} S. Langer {\it et al.},
Phys. Rev. B {\bf 79}, 214409 (2009).

\bibitem{karrasch2012} C. Karrasch, J. H. Bardarson, and J. E. Moore,
Phys. Rev. Lett. {\bf 108}, 227206 (2012).

\bibitem{karrasch2013} C. Karrasch {\it et al.},
Phys. Rev. B {\bf 87}, 245128 (2013).

\bibitem{karrasch2014} C. Karrasch, J. E. Moore, and F. Heidrich-Meisner,
Phys. Rev. B {\bf 89}, 075139 (2014).

\bibitem{znidaric2011} M. \v{Z}nidari\v{c},
Phys. Rev. Lett. {\bf 106}, 220601 (2011).

\bibitem{steinigeweg2012-2} R. Steinigeweg,
EPL {\bf 97}, 67001 (2012).

\bibitem{sirker2009} J. Sirker, R. G. Pereira, and I. Affleck,
Phys. Rev. Lett. {\bf 103}, 216602 (2009); Phys. Rev. B {\bf 83}, 035115 (2011).

\bibitem{fujimoto2003} S. Fujimoto and N. Kawakami,
Phys. Rev. Lett. {\bf 90}, 197202 (2003).

\bibitem{note}
For the values of $\Delta = 0.5$, $1.0$, $1.5$ considered here, and to the best
of our knowledge, we depict tDMRG data for the longest time scales available
from existing literature (Ref.~\cite{karrasch2012, karrasch2013, karrasch2014}).
This does {\it not} imply or suggest that longer times cannot be reached for
these or any other values of $\Delta$ using tDMRG.

\bibitem{gemmer2003} J. Gemmer and G. Mahler,
Eur. Phys. J. B {\bf 31}, 249 (2003).

\bibitem{goldstein2006} S. Goldstein {\it et al.},
Phys. Rev. Lett. {\bf 96}, 050403 (2006).

\bibitem{popescu2006} S. Popescu, A. J. Short, and A. Winter,
Nature Phys. {\bf 2}, 754 (2006).

\bibitem{reimann2007} P. Reimann,
Phys. Rev. Lett. {\bf 99}, 160404 (2007).

\bibitem{white2009} S. R. White,
Phys. Rev. Lett. {\bf 102}, 190601 (2009).

\bibitem{bartsch2009} C. Bartsch and J. Gemmer,
Phys. Rev. Lett. {\bf 102}, 110403 (2009); EPL {\bf 96}, 60008 (2011).

\bibitem{sugiura2012}
S. Sugiura and A. Shimizu,
Phys. Rev. Lett. {\bf 108}, 240401 (2012).

\bibitem{elsayed2013} T. A. Elsayed and B. V. Fine,
Phys. Rev. Lett. {\bf 110}, 070404 (2013).

\bibitem{steinigeweg2013} R. Steinigeweg {\it et al.},
arXiv:1311.0169 (2013), to appear in Phys. Rev. Lett.

\bibitem{deraedt2007} K. De Raedt {\it et al.},
Comp. Phys. Comm. {\bf 176}, 121 (2007).

\bibitem{jin2010} F. Jin {\it et al.},
J. Phys. Soc. Jpn. {\bf 79}, 124005 (2010).

\bibitem{karahalios2009} A. Karahalios {\it et al.},
Phys. Rev. B {\bf 79}, 024425 (2009).

\bibitem{chaloupka2010} J. Chaloupka, G. Jackeli, and G. Khaliullin,
Phys. Rev. Lett. {\bf 105}, 027204 (2010).

\end{thebibliography}
\end{document}